\documentclass[12pt, a4paper]{article}
\usepackage[utf8]{inputenc} % Se houver problema com acentuação de
\usepackage[english]{babel}
\usepackage[margin=2.5cm]{geometry}
\usepackage{setspace}
\usepackage{indentfirst}
\usepackage{graphicx}
\usepackage{xcolor}
\usepackage{times}
\usepackage{url}
\usepackage{enumerate}
\usepackage{amsmath, amsthm, amsfonts, amssymb, amsxtra}
\usepackage{bm}
\usepackage{natbib}
\usepackage{caption}
\usepackage{subcaption}

\newcommand{\bx}{\bm{x}}
\newcommand{\be}{\bm{e}}
\newcommand{\bfac}{\bm{f}}
\newcommand{\bnu}{\bm{\nu}}
\newcommand{\bepsilon}{\bm{\epsilon}}
\newcommand{\bchi}{\bm{\chi}}
\newcommand{\bC}{\bm{\mathrm{C}}}
\newcommand{\bLambda}{\bm{\Lambda}}
\newcommand{\bSigma}{\bm{\Sigma}}
\newcommand{\bTheta}{\bm{\Theta}}
\newcommand{\bPhi}{\bm{\Phi}}
\newcommand{\bGamma}{\bm{\Gamma}}

\allowdisplaybreaks

\begin{document}
\onehalfspacing

%=======================================================================
%============= MODIFIQUE O TEMPLATE A PARTIR DESTE PONTO ===============
%=======================================================================

% TÌTULO ------------------------------------

\begin{center}
  \textbf{\Large{A nonstationary seasonal Dynamic Factor Model: an application to temperature time series from the state of Minas Gerais}} \\[1em]
\end{center}

% AUTORES -----------------------------------

\begin{center}
  Davi Oliveira Chaves\footnote[1]{Institute of Mathematics and Statistics -- University of São Paulo (USP)}%
  \begingroup
    \renewcommand{\thefootnote}{\fnsymbol{footnote}}\footnote{Corresponding author: \textit{davi.chaves@ime.usp.br}}\endgroup,
  Chang Chiann\footnotemark[1], Pedro Alberto Morettin\footnotemark[1]
\end{center}

\vspace*{0.5cm}

% CORPO DO TRABALHO ----------------------------------------------------

\noindent In this paper, we apply a nonstationary seasonal Dynamic Factor Model to analyze temperature time serie from the state of Minas Gerais by extracting and studying their seasonal factors.

~


~

\noindent{\bf Acknowledgements:} The authors gratefully acknowledge financial support from the São Paulo Research Foundation (FAPESP) under project numbers 23/02538-0 and 23/11547-2.

\section{Introduction}

In many scientific fields, such as agriculture, temperature time series are of interest both as explanatory variables and as objects of study in their own right. However, at the state level, incorporating information from all possible locations in an analysis can be overwhelming, while using a summary measure, such as the state-wide average temperature, can result in significant information loss. In this context, using Dynamic Factor Models (DFMs) provides a compelling alternative for analyzing such multivariate time series, as they allow for the extraction of a small number of common factors that capture the majority of the variability in the data. Given that temperature series are typically seasonal, this paper applies the nonstationary seasonal DFM framework proposed by \cite{Nieto2016} to extract the common seasonal factors from monthly average temperature time series in the state of Minas Gerais (MG), Brazil. We then analyze these factors, along with their associated loadings, to gain a better understanding of the temperature dynamics in this state.

\section{Materials and Methods}

\subsection{Data}

\begin{figure}[htpb]
    % 800 x 350
    \centering
    \includegraphics[width=\textwidth]{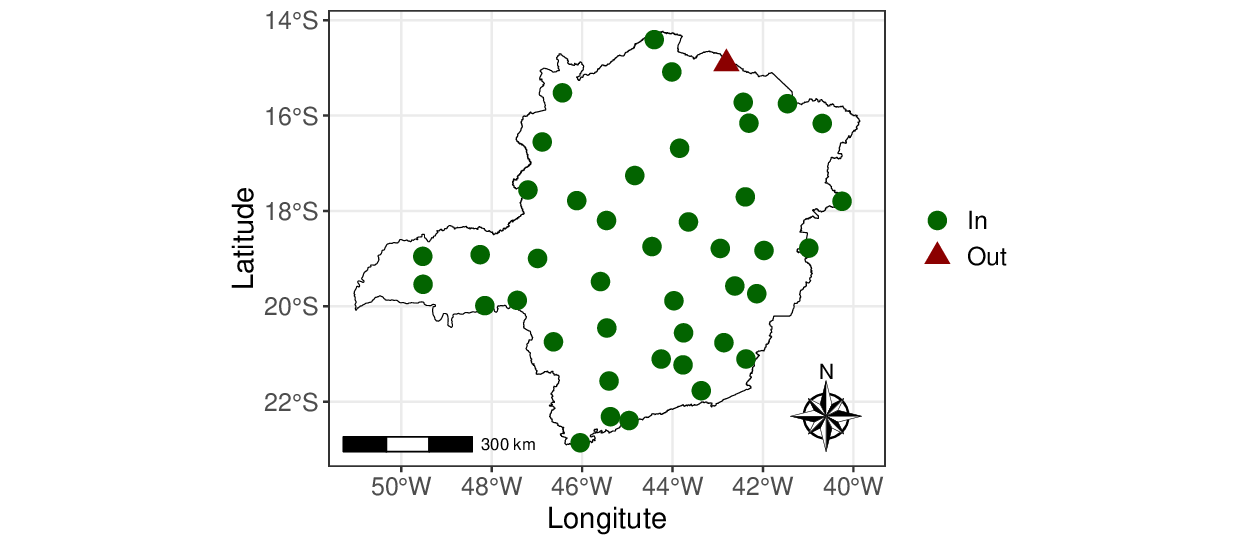}
    \caption{Map of automatic meteorological stations in MG that began operation before January 2008, with symbols indicating whether they were included in the analysis (green circles) or not (red triangles).}
    \label{fig:MapAllStations}
\end{figure}

In this study, we used monthly average temperatures (in degrees Celsius) from 42 meteorological stations in the state of MG, covering the period from January 2008 to December 2024, corresponding to a total of 204 time points. The data were obtained from the National Institute of Meteorology (INMET, from the Portuguese acronym). Initially, we selected all automatic meteorological stations in MG that began operation before January 2008, totaling 43 stations located in the following municipalities: Águas Vermelhas, Almenara, Barbacena, Belo Horizonte, Buritis, Camanducaia, Campina Verde, Capelinha, Caratinga, Conceição das Alagoas, Curvelo, Diamantina, Dores do Indaiá, Espinosa, Formiga, Governador Valadares, Guanhães, Guarda-Mor, Ituiutaba, Jaíba, João Pinheiro, Juiz de Fora, Mantena, Maria da Fé, Montalvânia, Montes Claros, Muriaé, Ouro Branco, Passa Quatro, Passos, Patrocínio, Pirapora, Rio Pardo de Minas, Sacramento, Salinas, São João del Rei, Serra dos Aimorés, Timóteo, Três Marias, Uberlândia, Unaí, Varginha, and Viçosa. As shown in Figure \ref{fig:MapAllStations}, these stations are well distributed across the state. One of them (those located in Espinosa), represented as the red triangle in the figure, exhibited an excessive amount of missing values and was excluded from the analysis, resulting in 42 stations.

Missing values in the remaining stations were imputed sequentially. For example, if the values for jan/13 and aug/20 were missing, the jan/13 value is imputed first, then, treating it as observed, the aug/20 value is subsequently imputed. 

Let $Y_t^{(1)} = \min\{5, \lfloor (t-1)/12 \rfloor\}$, and $Y_t^{(2)} = \min\{5, \lfloor (t-2)/12 \rfloor\}$, where $\lfloor\cdot\rfloor$ denotes the greatest integer less than or equal to a given number. The imputed value for a missing observation at time $t$, $\hat{x}_t$, is defined as
\begin{equation*}
    \hat{x}_t = \frac{\hat{x}_t^{(1)} + \hat{x}_t^{(2)}}{2},
\end{equation*}
where
\begin{equation*}
    \hat{x}_t^{(1)} = \frac{1}{Y_t^{(1)}} \sum_{i=1}^{Y_t^{(1)}} x_{t-12i} \quad \text{and} \quad \hat{x}_t^{(2)} = x_{t-1} + \frac{1}{Y_t^{(2)}} \sum_{i=1}^{Y_t^{(2)}} d_{t-12i}.
\end{equation*}
Here, $d_t$ denotes the diference between the value at time $t$ and the preceding one, that is, $d_t = x_t - x_{t-1}$.

In words, it means that the imputed value at time $t$ is given by the average between two estimates: (i) the mean of the values for the same month over the past $Y_t^{(1)}$ years, and (ii) the value of the previous month plus the mean of the differences between the value of the corresponding and its preceding months over the past $Y_t^{(2)}$ years. Here, $Y_t^{(1)}$ and $Y_t^{(2)}$ represent the minimum between $5$ and the number of past years available for the calculations ($\lfloor (t-1)/12 \rfloor$ or $\lfloor (t-2)/12 \rfloor$), as defined above. It means that the procedure aims to use the most recent five years of data to impute the current missing observations, nevertheless, if there are not five years available, all past years are used.

For illustration, consider the first missing value in jan/13 ($x_{61}$). Since 
\begin{equation*}
    Y_{61}^{(1)} = \min\{5, \lfloor (61-1)/12 \rfloor\} = 5,
\end{equation*}
the estimate $\hat{x}_{61}^{(1)}$ is computed using the values from the previous five Januaries: jan/12 ($x_{49}$), jan/11 ($x_{37}$), jan/10 ($x_{25}$), jan/09 ($x_{13}$), and jan/08 ($x_{1}$). Thus,
\begin{equation*}
    \hat{x}_{61}^{(1)} = \frac{1}{5}(x_{49} + x_{37} + x_{25} + x_{13} +x_{1}).
\end{equation*}
Similarly, since 
\begin{equation*}
    Y_{61}^{(2)} = \min\{5, \lfloor (61-2)/12 \rfloor\} = 4,
\end{equation*}
the estimate $\hat{x}_{61}^{(2)}$ is obtained using the most recent December (dec/12, $x_{60}$) and the four available December-January pairs: dec/11-jan/12 ($x_{48}:x_{49}$), dec/10-jan/11 ($x_{36}:x_{37}$), dec/09-jan/10 ($x_{24}:x_{25}$), and dec/08-jan/09 ($x_{12}:x_{13}$). Hence,
\begin{equation*}
    \hat{x}_{61}^{(2)} = x_{60} + \frac{1}{4}[(x_{49}-x_{48}) + (x_{37}-x_{36}) + (x_{25}-x_{24}) + (x_{13}-x_{12})].
\end{equation*}
Finally, the imputed value for $x_{61}$ is given by 
\begin{equation*}
    \hat{x}_{61} = \frac{\hat{x}_{61}^{(1)} + \hat{x}_{61}^{(2)}}{2}.
\end{equation*}

However, if the first missing values occur in the initial years of the series, there are none or very few past observations available for imputation. Therefore, for missing observations between jan/08 and jan/10, instead of using data from the previous five years, we used data from the following five years, even though some of these may also contain missing values. In such cases, the months with missing observations were excluded from the calculations.

Thus, the imputed value for a missing observation at time $1 \le t \le 25$, $\hat{x}_t$, is defined as
\begin{equation*}
    \hat{x}_t = \frac{\hat{x}_t^{(1)} + \hat{x}_t^{(2)}}{2},
\end{equation*}
where
\begin{equation*}
    \hat{x}_t^{(1)} = \frac{1}{Z_t^{(1)}} \sum_{i=1}^{5} x_{t+12i}^* \quad \text{and} \quad \hat{x}_t^{(2)} = x_{t-1} + \frac{1}{Z_t^{(2)}} \sum_{i=1}^{5} d_{t+12i}^*.
\end{equation*}
Here,
\begin{equation*}
    x_t^* = \begin{cases}
        0, \quad \text{if $x_t$ is missing}, \\
        x_t, \quad \text{otherwise},
    \end{cases} \quad \quad
    d_t^* = \begin{cases}
        0, \quad \text{if $x_t$ and/or $x_{t-1}$ is missing}, \\
        x_t - x_{t-1}, \quad \text{otherwise},
    \end{cases}
\end{equation*}
$Z_t^{(1)}$ denotes the number of non-missing values among $\{x_{t+12i}\}_{i=1}^5$, $i = 1, \dots, 5$, that are not missing, and $Z_t^{(2)}$ denotes the number of pairs $\{x_{t+12i}, x_{t+12i-1}\}_{i=1}^5$ for which neither value is missing.

For illustration, consider the first missing value in oct/08 ($x_{8}$). Also, assume that the values in oct/10 and sep/13 are missing. Then, $Z_8^{(1)} = 4$ and $Z_8^{(2)} = 3$. Since oct/10 is missing, the estimate $\hat{x}_{8}^{(1)}$ is computed using the values from the following four Octobers: oct/09 ($x_{20}$), oct/11 ($x_{44}$), oct/12 ($x_{56}$), and oct/13 ($x_{68}$). Thus,
\begin{equation*}
    \hat{x}_{8}^{(1)} = \frac{1}{4}(x_{20} + x_{44} + x_{56} + x_{68}).
\end{equation*}
Similarly, since oct/10 and sep/13 are missing, the estimate $\hat{x}_{8}^{(2)}$ is obtained using the most recent September (sep/08, $x_7$) and the three available September-October pairs in the next five years: sep/09-oct/09 ($x_{19}:x_{20}$), sep/11-oct/11 ($x_{43}:x_{44}$), and dec/12-jan/12 ($x_{55}:x_{56}$). Hence,
\begin{equation*}
    \hat{x}_{8}^{(2)} = x_{7} + \frac{1}{3}[(x_{20}-x_{19}) + (x_{44}-x_{43}) + (x_{56}-x_{55})].
\end{equation*}
Finally, the imputed value for $x_{8}$ is given by 
\begin{equation*}
    \hat{x}_{8} = \frac{\hat{x}_{8}^{(1)} + \hat{x}_{8}^{(2)}}{2}.
\end{equation*}

In summary, the imputed value for a missing observation at time $t$, $\hat{x}_t$, is defined as
\begin{equation*}
    \hat{x}_t = \frac{\hat{x}_t^{(1)} + \hat{x}_t^{(2)}}{2},
\end{equation*}
where
\begin{align*}
    & \hat{x}_t^{(1)} = \begin{cases}
        \frac{1}{Z_t^{(1)}} \sum_{i=1}^{5} x_{t+12i}^*, \quad \text{if} ~~ t \le 25, \\
        \frac{1}{Y_t^{(1)}} \sum_{i=1}^{Y_t^{(1)}} x_{t-12i}, \quad \text{if} ~~ t > 25, 
    \end{cases} \\
    & \hat{x}_t^{(2)} = \begin{cases}
        x_{t-1} + \frac{1}{Z_t^{(2)}} \sum_{i=1}^{5} d_{t+12i}^*, \quad \text{if} ~~ t \le 25, \\
        x_{t-1} + \frac{1}{Y_t^{(2)}} \sum_{i=1}^{Y_t^{(2)}} d_{t-12i}, \quad \text{if} ~~ t > 25.
    \end{cases}
\end{align*}

Among the $42$ variables, only six exhibited missing values between jan/08 and jan/10, and in all the cases there were sufficient data in the subsequent five years to perform imputation. It is also worth noting that if the first observation, $x_1$, is missing, the estimate $\hat{x}_1^{(2)}$ cannot be computed. A possible solution would be to rely solely on $\hat{x}_1^{(1)}$ for imputation or to further complicate the methodoly by designing a special imputation procedure for this case. Fortunately, this observation is available for all variables in our dataset.

After imputation, the data were standardized, meaning that each variable was centered by subtracting its mean and scaled by dividing by its standard deviation, before the DFM analysis. This step was performed to mitigate scale effects, as the annual temperature range can vary substantially across locations.

\subsection{Theoretical Background}

The centered (zero-mean) temperature time series under study is modeled as a nonstationary DFM given by
\begin{equation*}
    \bx_t = \bLambda_1 \bfac_{1t} + \bLambda_2 \bfac_{2t} + \bLambda_3 \bfac_{3t} + \bepsilon_t = \bLambda \bfac_t + \bepsilon_t, \quad t = 1, \dots, T,
\end{equation*}
where $\bx_t$ is the $n$-dimensional zero-mean observed multivariate time series, $\bLambda = [\bLambda_1 ~ \bLambda_2 ~ \bLambda_3]$ is the $n \times r$ matrix of loadings, and $\bfac_t = (\bfac_{1t}', \bfac_{2t}', \bfac_{3t}')'$ is the $r$-dimensional vector of latent factors, with $r \ll n$. The $n$-dimensional vector $\bchi_t = \bLambda \bfac_t$ is referred to as the common component. When the factors capture a large proportion of the variability of the original series, $\bchi_t$ closely approximates $\bx_t$. The term $\bepsilon_t$ denotes an $n$-dimensional Gaussian white noise process with mean $\bm{0}_n$ and diagonal covariance matrix $\bSigma_\epsilon$, it is known as the idiosyncratic component and accounts for dynamics that are specific to each individual series.

The factor vector $\bfac_t$ consists of three distinct processes: the first, $\bfac_{1t}$, of dimension $r_1$, is nonstationary and nonseasonal, and follows the model
\begin{equation*}
    \bPhi_1(B)(1-B)^d \bfac_{1t} = \bTheta_1(B) \bnu_{1t},
\end{equation*}
with $d \ge 1$; the second, $\bfac_{2t}$, of dimension $r_2$, is nonstationary and seasonal, and follows the model
\begin{equation*}
    \bPhi_2(B^S)(1-B^S)^D \bfac_{2t} = \bTheta_2(B^S) \bnu_{2t},
\end{equation*}
with $D \ge 1$ and $S$ representing the seasonal period; the third and final component, $\bfac_{3t}$, of dimension $r_3$, is stationary, and follows the model
\begin{equation*}
    \bPhi_3(B) \bfac_{3t} = \bTheta_3(B) \bnu_{3t}.
\end{equation*}

For all three process, the following assumptions hold: $\bnu_{it}$ is an $r_i$-dimensional Gaussian white noise process with mean $\bm{0}_{r_i}$ and diagonal covariance matrix $\bSigma_{\nu_i}$; the processes $\bnu_{it}$ and $\bnu_{jt}$ are independent for $i \ne j$; the factor innovation vector $\bnu_t = (\bnu_{1t}', \bnu_{2t}', \bnu_{3t}')'$ is independent of $\bepsilon_s$ for all $t, s \in \{1, \dots, T\}$; the matrix polynomial operators $\bPhi_i(B)$ and $\bTheta_i(B)$, for $i = 1,2, 3$, are diagonal, and their diagonal elements satisfy the usual conditions to ensure that the processes $(1-B)^d \bfac_{1t}$, $(1-B^S)^D \bfac_{2t}$, and $\bfac_{3t}$ are stable and invertible.

This model was introduced by \cite{Nieto2016}. Assuming for simplicity that $d = D$, it was shown therein that the sample generalized autocovariance (SGCV) matrix 
\begin{equation}
    \label{eq:SGCV}
    \bC(h) = (S/T)^{2d} \sum_{t=h+1}^T \bx_t \bx_{t-h}',
\end{equation}
where $h$ denotes the lag, converges weakly, as $T \to \infty$ with $h$ fixed, to a random matrix $\bGamma(h)$. This limiting matrix $\bGamma(h)$ almost surely has $r_1 + r_2$ positive eigenvalues and $n - r_1 - r_2$ zero eigenvalues if $h$ is a multiple of $S$, and $r_1$ positive eigenvalues and $n - r_1$ zero eigenvalues otherwise. This property allows $\bC(h)$ to be used for identifying the number of factors in the DFM.

The nonstationary factors, $\bfac_{1t}$ and $\bfac_{2t}$, along with their corresponding loadings, $\bLambda_1$ and $\bLambda_2$, can be estimated via Principal Component Analysis (PCA) using the SGCV matrix. Specifically, the nonstationary loading $\bLambda^{(ns)} = [\bLambda_1 ~ \bLambda_2]$ is estimated as the orthonormal eigenvectors corresponding to the $r_1 + r_2$ largest eigenvalues of the matrix $\bC(h)$, with $h$ being a multiple of $S$ (e.g, $h = S$). Then, the nonstationary factors $\bfac_t^{(ns)} = (\bfac_{1t}', \bfac_{2t}')'$ are estimated as
\begin{equation*}
    \hat{\bfac}_t^{(ns)} = [\hat{\bLambda}^{(ns)}]' \bx_t, \quad t = 1, \dots, T.
\end{equation*}
Subsequently, PCA is applied to the residuals $\hat{\be}_t^{(ns)} = \bx_t - \hat{\bLambda}^{(ns)} \hat{\bfac}_t^{(ns)}$ using their sample covariance matrix to estimate the stationary factor $\bfac_{3t}$ and its corresponding loading $\bLambda_3$.

\section{Results and Discussion}

\begin{figure}[htbp]
    % 800 x 250
    \centering
    \includegraphics[width=\textwidth]{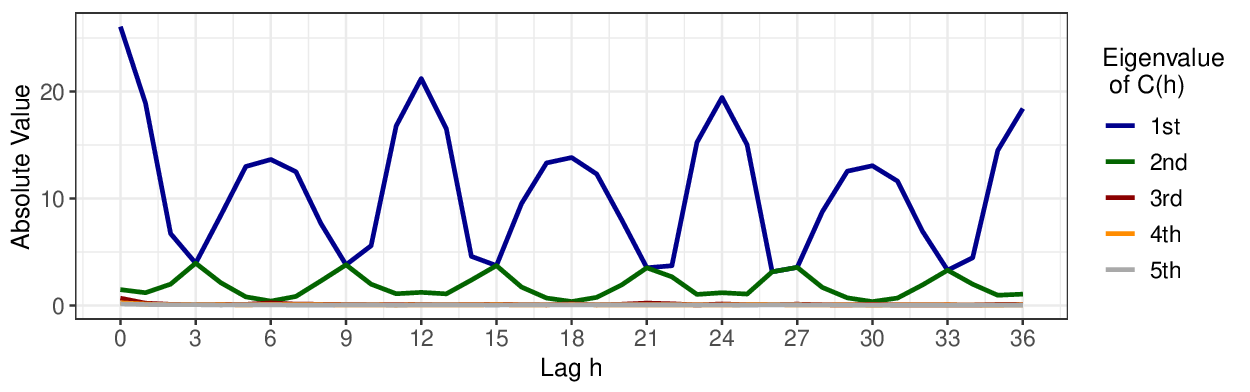}
    \caption{Sequence of absolute values of the five largest eigenvalues of $\bC(h)$ from lag $h = 0$ to $h = 36$.}
    \label{fig:Eigenvalues}
\end{figure}

First, it is necessary to determine the number of factors. To this end, we analyzed the sequence of eigenvalues of the SGCV matrix defined in \eqref{eq:SGCV}, for lags ranging from $0$ to $36$, as depicted in Figure \ref{fig:Eigenvalues}. The figure reveals two clearly distinguishable nonzero eigenvalues. The spikes of the first eigevalue align with the seasonal lags and indicate the presence of a seasonal factor. The spikes of the second eigenvalue, although not occurring precisely at seasonal lags, emerge between those of the first and exhibit a cyclical pattern, suggesting that it also represents a seasonal factor. Indeed, if a DFM is estimated with only one seasonal factor and this eigenvalue analysis is repeated on the residuals, the first eigenvalue (which in this case corresponds to the second one from the original data) then displays spikes at the seasonal lags. Based on this evidence, we choose to fit a DFM with two seasonal factors ($r_2 = 2$), no nonstationary nonseasonal factors ($r_1 = 0$), and no stationary factors ($r_3 = 0$).

The estimated loading matrix, $\hat{\bLambda}$, was obtained from the orthonormal eigenvectors corresponding to the two largest eigenvalues of $\bC(12)$. The factors are then estimated as
\begin{equation*}
    \hat{\bfac}_t = \hat{\bLambda}' \bx_t, \quad t = 1, \dots, T.
\end{equation*}
Finally, the estimated common components are computed as
\begin{equation*}
    \hat{\bchi}_t = \hat{\bLambda} \hat{\bfac}_t, \quad t = 1, \dots, T.
\end{equation*}

\begin{figure}[htbp]
    % 800 x 250
    \centering
    \begin{subfigure}[b]{0.99\textwidth}
        \centering
        \includegraphics[width=\textwidth]{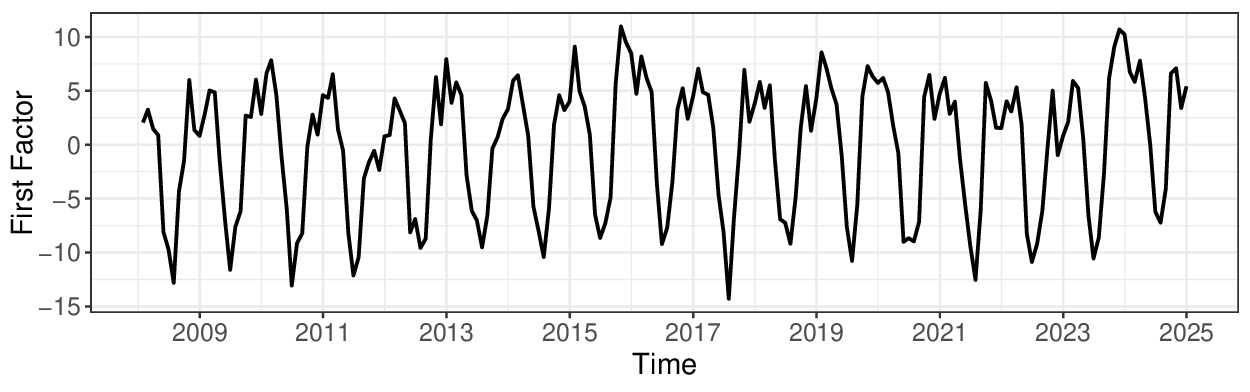}
    \end{subfigure}
    \vspace{1em}
    \begin{subfigure}[b]{0.99\textwidth}
        \centering
        \includegraphics[width=\textwidth]{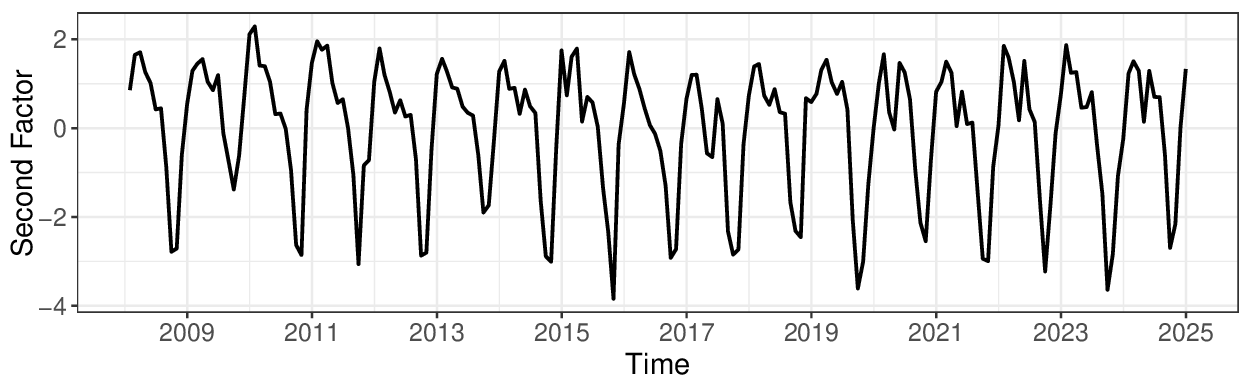}
    \end{subfigure}
    \caption{The two seasonal factors.}
    \label{fig:Factors}
\end{figure}

\begin{figure}[htbp]
    % 450 x 550
    \centering
    \begin{subfigure}[b]{0.48\textwidth}
        \centering
        \includegraphics[width=\textwidth]{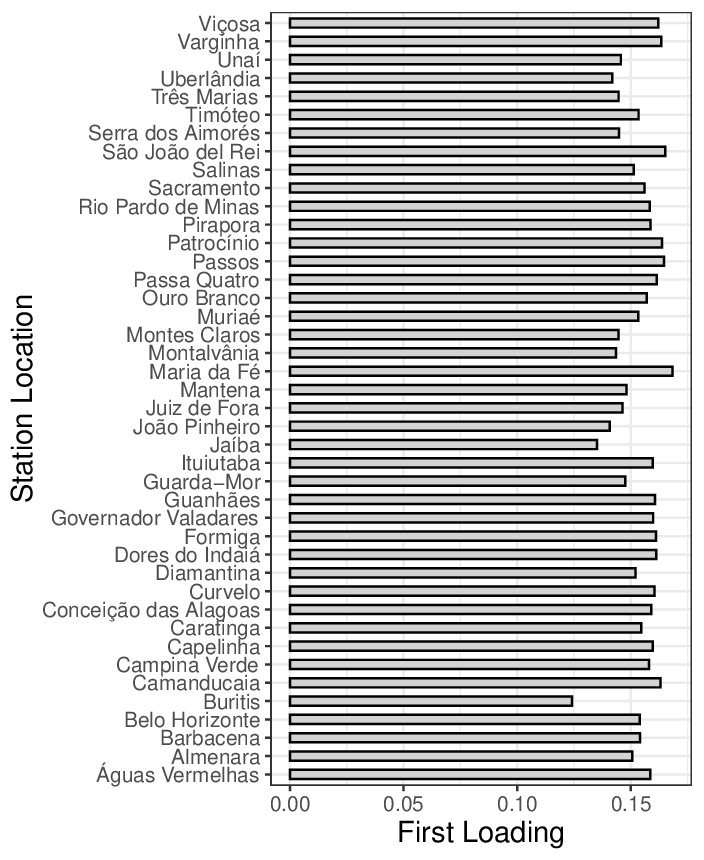}
    \end{subfigure}
    \hfill
    \begin{subfigure}[b]{0.48\textwidth}
        \centering
        \includegraphics[width=\textwidth]{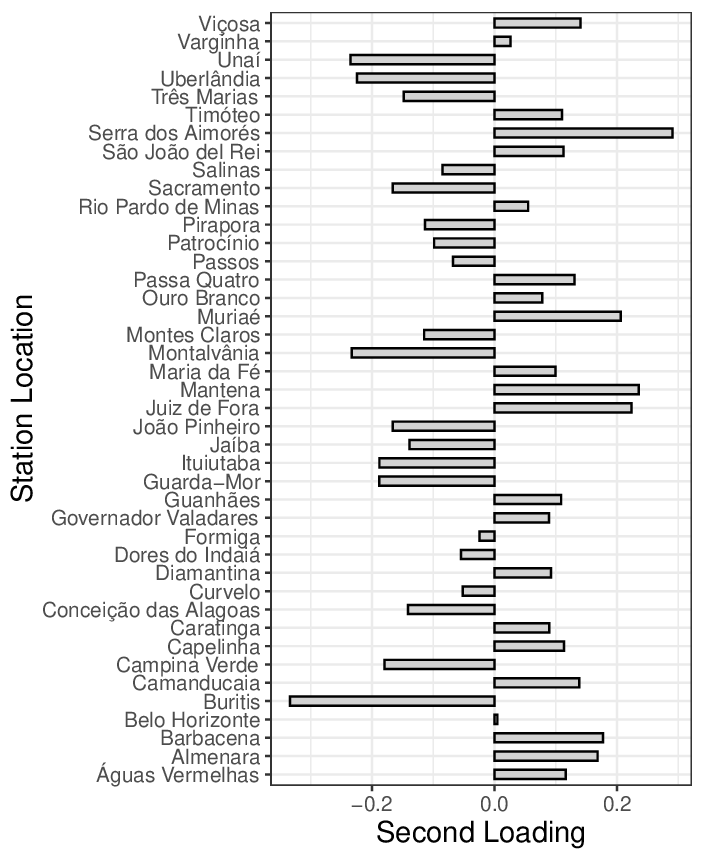}
    \end{subfigure}
    \caption{Loadings corresponding to the two factors, displayed by the municipality where the meteorological station is located.}
    \label{fig:Loadings}
\end{figure}

The estimated factors and loadings are shown in Figures \ref{fig:Factors} and \ref{fig:Loadings}, respectively. The presence of seasonality is clearly visible in both factors. Furthermore, by examining the loadings, it becomes evident that the first factor captures the overall seasonal pattern across MG, while the second factor highlights contrasting seasonal behaviors between two distinct groups of locations within the state.

\begin{figure}[htpb]
    % 800 x 300
    \centering
    \begin{subfigure}[b]{0.99\textwidth}
        \centering
        \includegraphics[width=\textwidth]{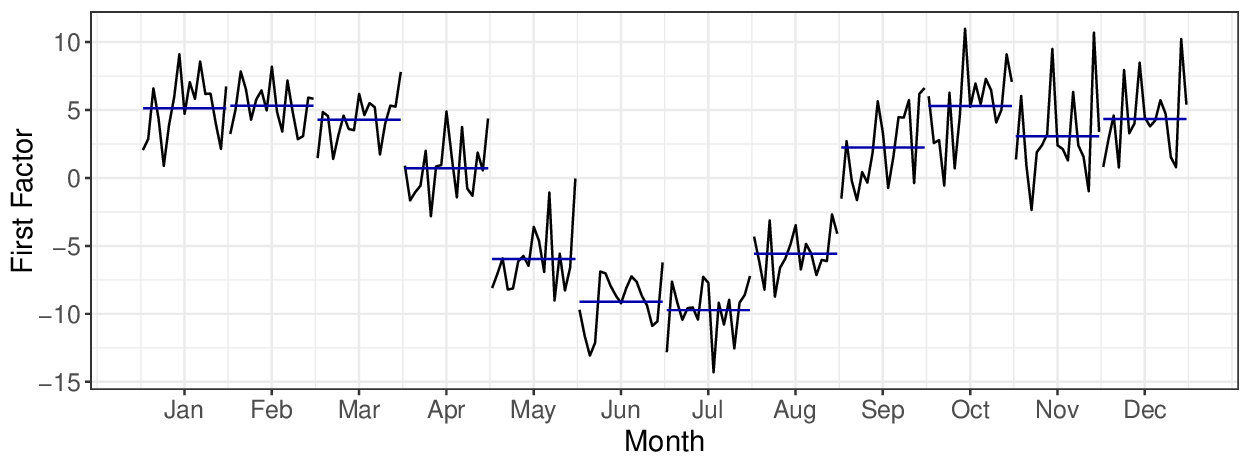}
    \end{subfigure}
    \vspace{1em}
    \begin{subfigure}[b]{0.99\textwidth}
        \centering
        \includegraphics[width=\textwidth]{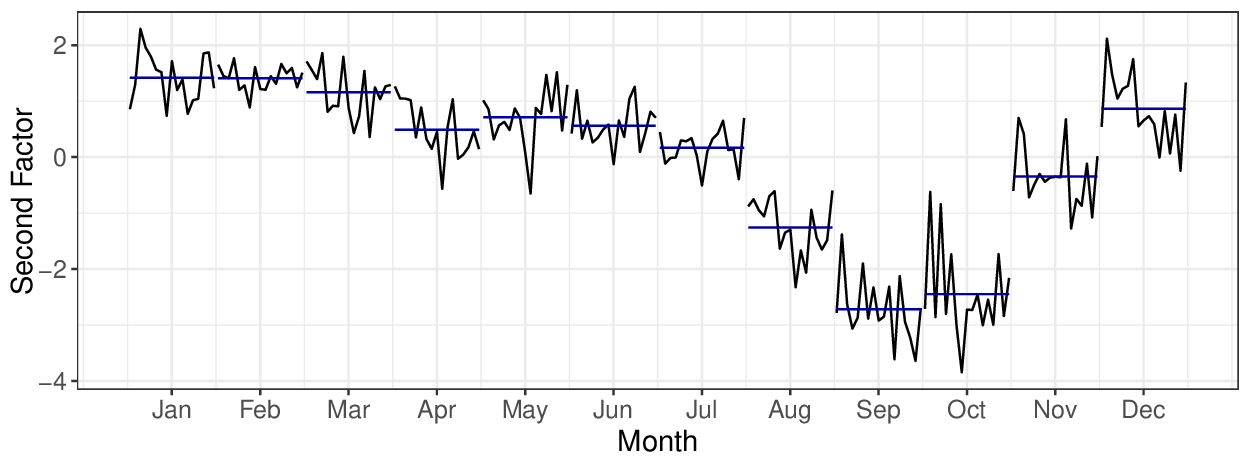}
    \end{subfigure}
    \caption{The factor evolution for each month together with its mean (blue solid line), summarizing the annual seasonal pattern of the two factors.}
    \label{fig:FactorsSeasonal}
\end{figure}

The first factor is well modeled by a SARIMA$(2,0,0)(0,1,1)_{12}$ process, and its seasonal pattern is shown at the top of Figure \ref{fig:FactorsSeasonal}. This pattern reflects the typical climate behavior of the state, with lower temperatures during the fall and winter, and higher temperatures in spring and summer. Additionaly, at the top of Figure \ref{fig:Factors}, a modest upward trend can be observed toward the end of the series, which may be associated with global warming.

\begin{figure}[htpb]
    \centering
    \includegraphics[width=\textwidth]{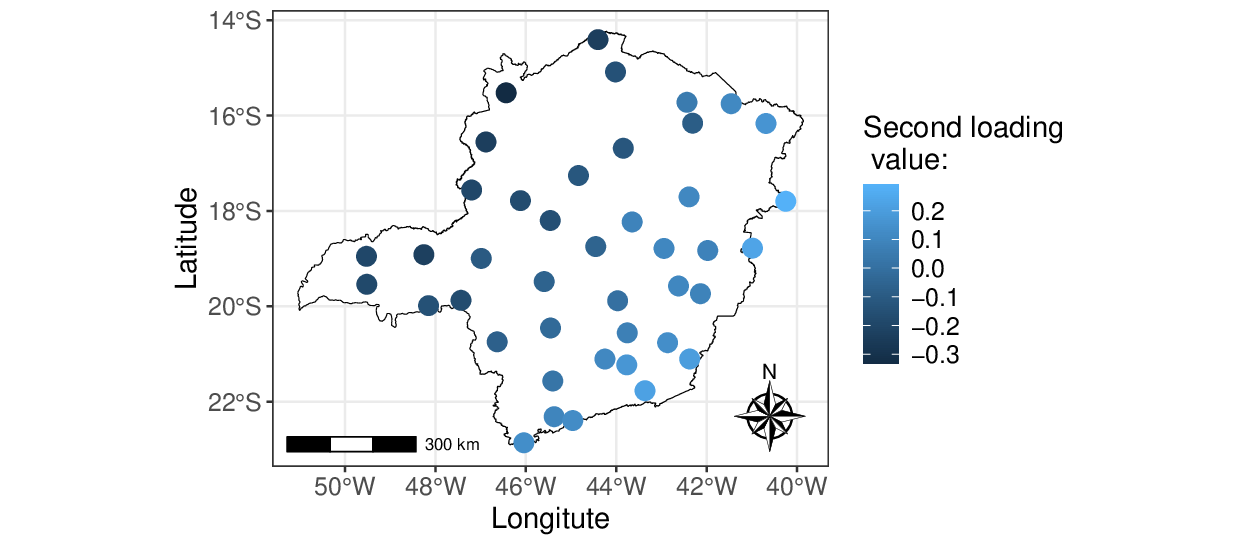}
    \caption{Map of meteorological stations in Minas Gerais, colored and shaped according to the second loading value.}
    \label{fig:LoadingMap}
\end{figure}

The second factor is well modeled by a SARIMA$(1,0,0)(0,1,1)_{12}$ process, and its seasonal pattern is shown at the bottom of Figure \ref{fig:FactorsSeasonal}. It exhibits lower (and negative) values during the early spring months, September and October, and higher (and positive) values in the summer months, December, January and February. This pattern, combined with the map of stations colored according to the second loading values shown in Figure \ref{fig:LoadingMap}, highlights a contrast between the seasonal behaviors of two distinct regions in MG. The stations shown in lighter blue, primarily located in the Atlantic Forest biome, are positively associated with the second factor. This implies that, adding this to the general seasonal pattern captured by the first factor results in higher annual temperatures in summer for these places. In contrast, the stations depicted in darker blue, predominantly situated in the Cerrado (Brazilian Savanna) biome, are negatively associated with the second factor. As a result, they usually exhibit higher annual temperatures in spring, particularly in September and October.

\begin{figure}[htpb]
    % 800 x 300
    \centering
    \begin{subfigure}[b]{0.99\textwidth}
        \centering
        \includegraphics[width=\textwidth]{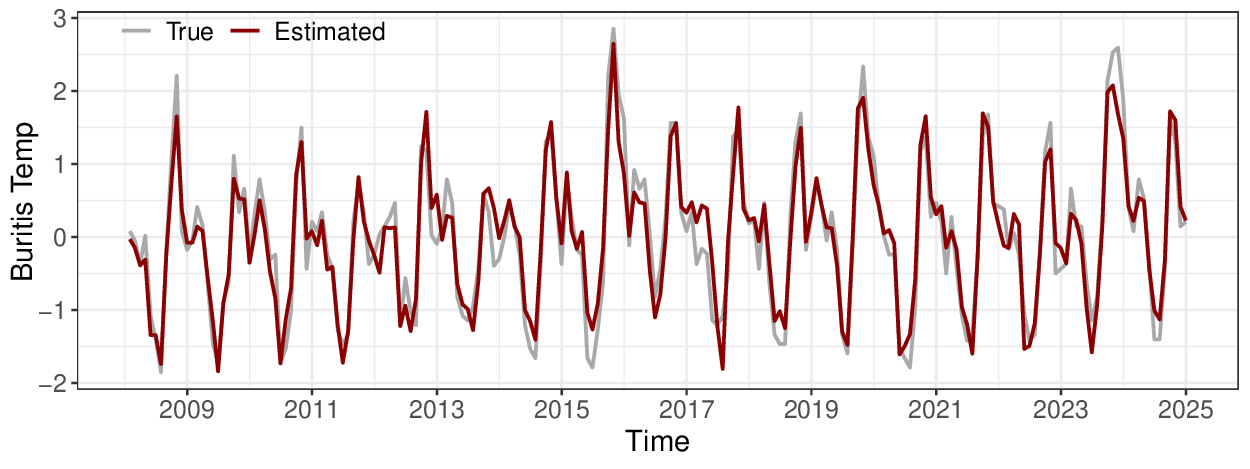}
    \end{subfigure}
    \vspace{1em}
    \begin{subfigure}[b]{0.99\textwidth}
        \centering
        \includegraphics[width=\textwidth]{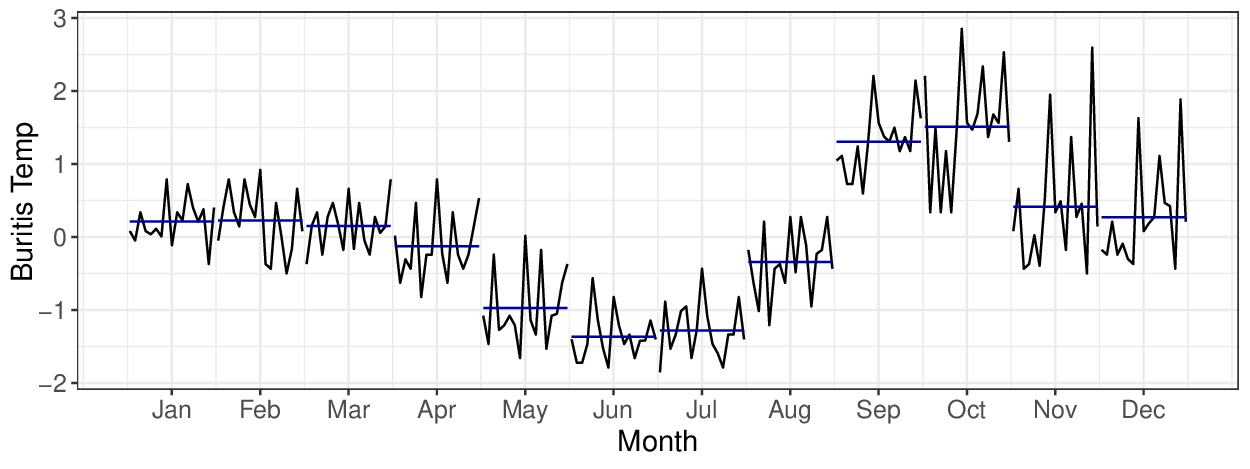}
    \end{subfigure}
    \caption{At the top, the standardized temperature time series for Buritis (grey line) is shown together with the corresponding estimated common component (red line). At the bottom, the monthly time series are displayed together with their mean (blue solid line), summarizing the annual seasonal temperature pattern in Buritis.}
    \label{fig:Buritis}
\end{figure}

\begin{figure}[htpb]
    % 800 x 300
    \centering
    \begin{subfigure}[b]{0.99\textwidth}
        \centering
        \includegraphics[width=\textwidth]{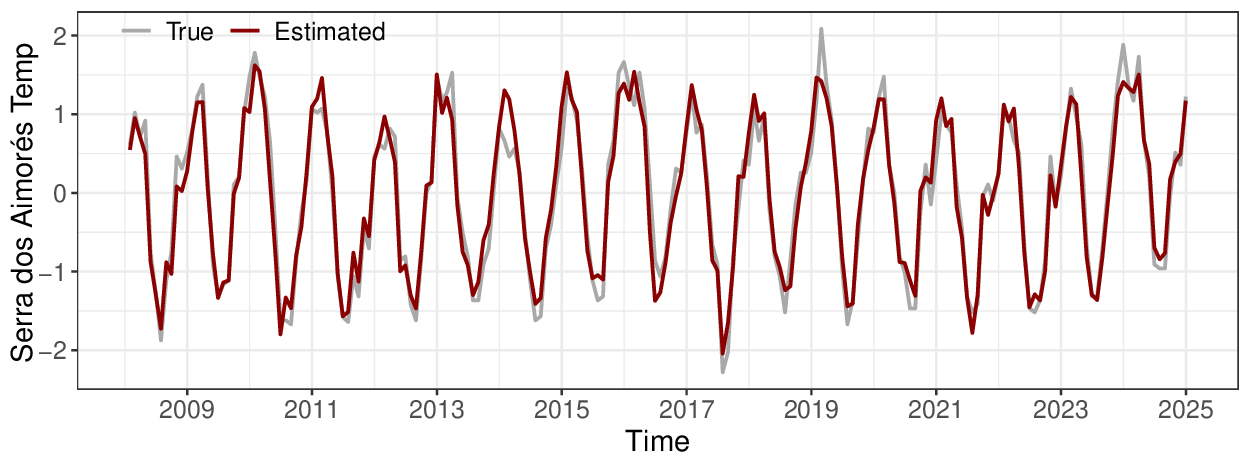}
    \end{subfigure}
    \vspace{1em}
    \begin{subfigure}[b]{0.99\textwidth}
        \centering
        \includegraphics[width=\textwidth]{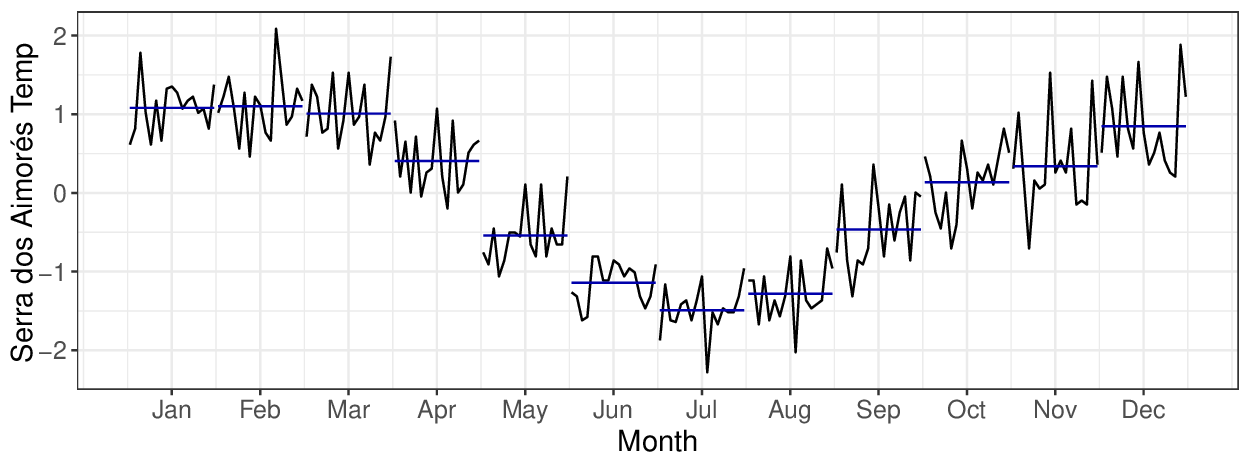}
    \end{subfigure}
    \caption{At the top, the standardized temperature time series for Serra dos Aimorés (grey line) is shown together with the corresponding estimated common component (red line). At the bottom, the monthly time series are displayed together with their mean (blue solid line), summarizing the annual seasonal temperature pattern in Serra dos Aimorés.}
    \label{fig:SerraAimores}
\end{figure}

To illustrate how effectively these two estimated factors capture the variability of the dataset under study, the top panels of Figures \ref{fig:Buritis} and \ref{fig:SerraAimores} allow a visual comparison between the observed temperature values (grey lines) at two stations, located in Buritis and Serra dos Aimorés, and the corresponding estimated common components (red lines). Notably, the estimated common components closely approximate the observed values in these examples. A similar pattern emerges across the other stations, providing strong evidence that the two estimated factors adequately represent the variability contained in the $42$ time series.

These two illustrative stations were not chosen arbitrarily. As shown in Figure \ref{fig:Loadings}, they exhibit the largest absolute values of the second loading, with Buritis being negative and Serra dos Aimorés positive. Consequently, in Figure \ref{fig:LoadingMap}, Buritis appears as the darkest dot, located at the northwestern extremity of the state, while Serra dos Aimorés appears as the lighest dot, located farthest to the east. This implies that, according to what was previously discussed, Buritis should exhibit higher annual temperatures during spring, particularly in September and October, whereas Serra dos Aimorés should exhibit them during summer. That pattern is indeed observed and can be confirmed in the bottom panels of Figures \ref{fig:Buritis} and \ref{fig:SerraAimores}.

\section{Conclusion}

In conclusion, we saw that the $42$-dimensional temperature time series from MG can be effectively represented using only two seasonal factors. The first one captures the general seasonal pattern of the state, while the second contrasts the months of higher annual temperatures between two distinct regions: the inland region closer to the center of Brazil, which exhibits peak temperatures in Septermber and October, and the region closer to the coast, which experiences higher temperatures during the summer months.

\bibliography{biblio}

\end{document}